\documentclass[prl,twocolumn,showpacs]{revtex4}
\usepackage{graphicx}
\usepackage{amssymb}
\usepackage{color}

\newcommand{\be}{\begin{equation}}
\newcommand{\ee}{\end{equation}}
\newcommand{\bea}{\begin{eqnarray}}
\newcommand{\eea}{\end{eqnarray}}
\newcommand{\pf}{}
\newcommand{\mv}{}
\newcommand{\Zed}{{\cal Z}}
\begin{document}

\title{Quantum criticality in a double quantum-dot system}
\author{
Gergely Zar\'and,$^{1,2}$ Chung-Hou Chung,$^{3}$ Pascal Simon,$^{4}$ and
Matthias Vojta$^{3}$
}
\affiliation{
\mbox{$^1$ Institut f\"{u}r Theoretische Festk\"{o}rperphysik, Universit\"{a}t Karlsruhe, D-76128 Karlsruhe,
Germany}
\\
$^2$ Institute of Physics, Technical University Budapest, Budapest, H-1521, Hungary
\\
\mbox{$^3$ Institut f\"ur Theorie der Kondensierten Materie, Universit\"{a}t Karlsruhe, D-76128 Karlsruhe,
Germany}
\\
$^4$Laboratoire de Physique et Mod\'elisation des Milieux Condens\'es, CNRS et Universit\'e
Joseph Fourier, 38042 Grenoble, France
}
\date{Sep 6, 2006}

\begin{abstract}
We discuss the realization of the quantum-critical non-Fermi liquid state,
originally discovered within the two-impurity Kondo model,
in double quantum-dot systems.
Contrary to the common belief, the corresponding fixed point is robust against
particle-hole and various other asymmetries, and is only unstable to charge transfer
between the two dots.
We propose an experimental set-up where such charge transfer processes are suppressed,
allowing a controlled approach to the quantum critical state.
We also discuss transport and scaling properties in the vicinity of the critical point.
\end{abstract}

\maketitle


Quantum dots can be used to build single-electron transistors \cite{SET}
and spin-based quantum bits \cite{Marcus},
but equally interestingly, they serve as artificial atoms and allow to
access correlated states of matter \cite{David,ST,SU4}.
So far, most experiments focused on the study of Fermi-liquid states,
with regular thermodynamic and transport properties at low temperatures \cite{David,SU4}
and simple transitions or crossovers between them \cite{ST}.
However, artificial molecules and mesoscopic structures can be used to realize and
study non-Fermi liquids as well,
characterized by singular properties and providing the simplest examples of
quantum critical systems.
However, due to their singular nature, these states are very elusive.
In fact, only recently Oreg and Goldhaber-Gordon \cite{Oreg} proposed a controlled
set-up to access the two-channel Kondo (2CK) fixed point \cite{Cox,matveev},
being the paradigmatic example of non-Fermi liquid impurity system.
Subsequently, this setup was successfully realized experimentally \cite{exp2ck}.
{\mv
Dissipation has also been proposed to drive quantum phase transitions (QPT)
in quantum dots \cite{LeHur,BSGG}.
However, most dissipative QPT are of Kosterlitz-Thouless type,
and therefore no true quantum-critical state is realized.}

A non-Fermi-liquid state, similar to the one of the 2CK model,
emerges in the two-impurity Kondo model (2IKM).
This model, initially studied in the context of heavy-fermion QPT,
consists of two impurity spins that are coupled to conduction electrons and, at the same
time, interact with each other through an exchange interaction.
Jones {\em et al.} \cite{JonesVarmaWilkins1} observed that in the 2IKM
a quantum critical point (QCP) separates a ``local-singlet'' from a Kondo-screened phase.
This QCP has been shown to be essentially equivalent to the 2CK fixed
point \cite{Gan}, though its operator content and finite-size spectrum
are different \cite{ALJ}.
In fact, it has been observed that -- unlike the 2CK fixed point --
the QCP of the 2IKM is very sensitive to certain electron-hole symmetry-breaking
processes, which can smooth the QPT into a cross-over \cite{ALJ,Sakai}.
(A related non-Fermi liquid fixed point also appeared in a two-orbital Anderson model
\cite{Fabrizio}.)

\begin{figure}[b]
\includegraphics[width=7cm]{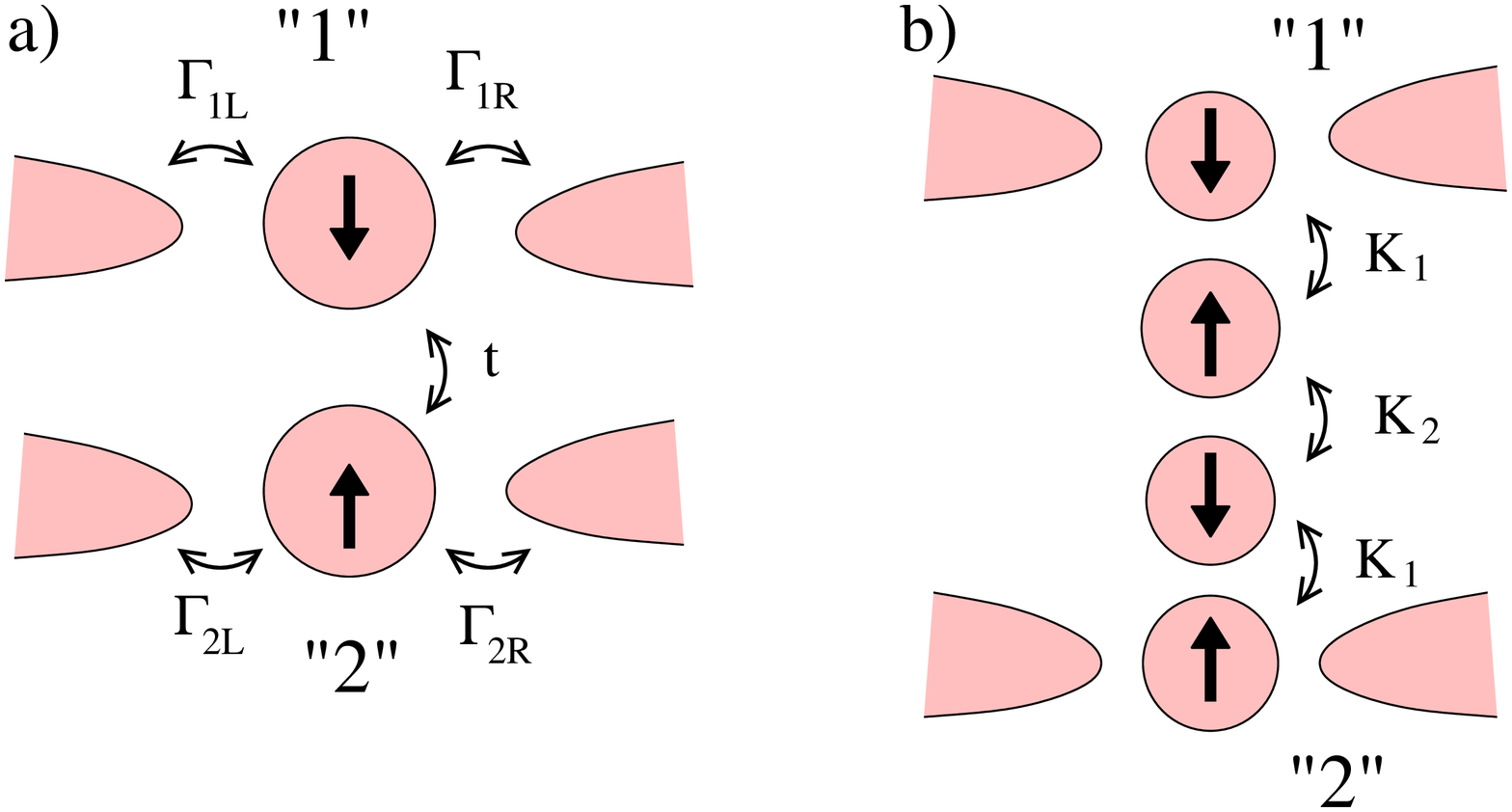}
\caption{\label{fig:DD}
a) System of two quantum dots studied in the paper.
b) Modified set-up with suppressed charge transfer processes, with an even number of quantum
dots inserted between the two main dots (1,2) attached to leads.
}
\vspace*{-5pt}
\end{figure}

The purpose of the present paper is to demonstrate that the QCP of the 2IKM
can be realized and studied in a system of two quantum dots,
shown in Fig.~\ref{fig:DD}.
Such a double-dot system has a number of interesting regimes \cite{GlazmanDD},
however, here we shall focus on a situation far from the charge degeneracy points,
with one unpaired electron on each of the dots.
Remarkably, the quantum critical state in this geometry is
very robust against both the asymmetry of the device (parity) and
electron-hole asymmetry, and a sharp phase transition appears
as long as there is {\em no charge transfer} between the dots 1 and 2.
We show that these charge transfer processes
can be suppressed by inserting an artificial ``antiferromagnetic insulator''
between the two dots (see Fig.~\ref{fig:DD}b).

{\em Model.}
To start our analysis, let us assume that the charging energies $E_{C1,2}$
($E_{C1} \approx E_{C2}\approx E_C$), associated with putting an extra electron
to one of the two dots, are large compared to the level widths of the dots,
$\Gamma_\gamma$ ($\gamma=1,2$), and to the tunneling $t$ between the two dots.
Perturbatively integrating out virtual charge fluctuations of the dots,
we arrive at the following Hamiltonian:
\bea
 H_{\rm int} & = & K \; \vec S_1\cdot \vec S_2 +
 \frac 12 (J^{(1)}_{\gamma\gamma'}  \vec S_1 + J^{(2)}_{\gamma\gamma'}  \vec S_2)
 \psi^\dagger_{\gamma} \vec \sigma \psi_{\gamma'}
\label{eq:H_gen}
\\
 &+& ( V_{\gamma\gamma'}+ Q_{\gamma\gamma'}  \vec S_1\cdot \vec S_2)
 \psi^\dagger_{\gamma} \psi_{\gamma'} + \mbox{irrelevant terms}.
\nonumber \eea
Here $\psi^\dagger_{\gamma \sigma } = \varrho_\gamma^{1/2} \sum_\epsilon
c^\dagger_{\epsilon\gamma\sigma}$, with $c^\dagger_{\epsilon\gamma\sigma}$ being
the creation operator of an electron state with spin $\sigma$ and energy $\epsilon$ in
the even combination of electrons in the leads attached to dot $\gamma$, and $\varrho_\gamma$
their density of states at the Fermi energy.
Apart from irrelevant terms, Eq.~(\ref{eq:H_gen}) is the most general Hamiltonian
that describes the double-dot system in the regime where charge fluctuations are suppressed.
The largest couplings are $K$,
$J_\gamma \equiv J^{(\gamma)}_{\gamma\gamma}$, and
$V_\gamma \equiv V_{\gamma\gamma}$,
since these couplings are generated by second-order tunneling processes.
They are typically of the size $J_1 \sim V_1 \sim \Gamma_1/E_{C}$,
$J_2 \sim V_2 \sim \Gamma_2/E_{C}$, and $K \sim t^2/E_{C}$.
The couplings $V_1$ and $V_2$ can be made small by tuning the dots close to the middle of
their respective Coulomb blockade valleys.
The second-largest couplings are associated with {\em charge transfer}
between leads 1 and 2, and are all of order
$V_{12} \sim Q_{12} \sim J^{(\gamma)}_{12} \sim (J_1 J_2 K/E_C)^{1/2}$.
All other couplings are suppressed by further powers of $t/E_C$, $\Gamma/E_C$,
and do not change the physics essentially.

Let us first study the Hamiltonian with the leading terms only,
and no charge transfer between the two sides:
\bea
 \tilde H_{\rm int} & = & K \; \vec S_1 \vec S_2 +
 \frac 12 (
J_1  \vec S_1  \psi^\dagger_{1} \vec \sigma \psi_{1}
+J_2  \vec S_2  \psi^\dagger_{2} \vec \sigma \psi_{2}
)
\label{eq:H_simple}
\\
&+&  V_1 \psi^\dagger_{1} \psi_{1}  + V_2 \psi^\dagger_{2} \psi_{2}\;.
\nonumber
\eea
This Hamiltonian is characterized by three energy scales:
Without the coupling $K$, the two spins on the two dots
are screened independently at the Kondo temperatures
$T_1 \approx  \delta\epsilon \;e^{-1/J_1}$ and $T_2 \approx  \delta\epsilon \;e^{-1/J_2}$,
with  $\delta\epsilon\ll E_C$  the typical level spacing on
the dots \cite{renormfoot}.
These Kondo scales compete with $K$ that tends to bind the two spins into an
inter-impurity singlet.

Clearly, the terms in Eq.~(\ref{eq:H_simple}) may break both parity and electron-hole
symmetry.
Nevertheless, solving Eq.~(\ref{eq:H_simple}) using a
numerical renormalization group (NRG) approach
we find a sharp QPT upon variation of $K$
for {\em any} value of the couplings $J_\gamma$ and $V_\gamma$
{\mv (in contrast to earlier statements)}.
In all cases, the spectrum at the critical point can be described through
a generalized version of the conformal field theory (CFT)
of Affleck {\em et al.}~\cite{ALJ}, to be discussed below.

\begin{figure}
\includegraphics[width=8cm]{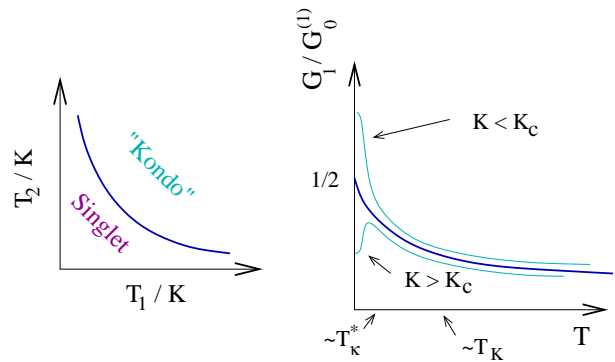}
\caption{\label{fig:phase}
Left: Phase diagram of the double-dot device in the absence of charge
transfer. The two phases are separated by line of second order QPT,
being very similar to the two-channel Kondo state.
Right: Sketch of the temperature-dependent conductance through dot "1" for
$T_1\approx T_2\approx T_K$, in the absence of charge transfer between the two sides.
}
\vspace*{-5pt}
\end{figure}

{\em Asymmetric limit.}
Before diving into the CFT solution, let us give a simple and revealing
physical picture of the physics in the limit $T_1 \gg K \gg T_2$.
Here, the first spin is screened at a temperature $T\sim T_1$.
Below that scale, a local Fermi-liquid description applies to the resulting
Kondo-screened complex, and therefore, it acts as a {\em bath} which tries to screen
the spin $S_2$ \cite{two-stage}.
The effective dimensionless coupling between $S_2$ and the Kondo complex can be estimated as
$\lambda_{1}\sim K/T_1$. However, $S_2$ also couples to spin excitations
in the leads attached to it, with a renormalized coupling $\lambda_2 \approx 1/\ln(T_1/T_2)$.
Clearly, we end up with an effective 2CK model, which is known to display a QPT at
$\lambda_1 = \lambda_2$, corresponding to the condition
$T_2 \approx T_1 \exp(- a \; T_1/K)$,
with $a$ a constant of the order of unity.
The above argument is independent of the potential scattering terms.
It shows that
(i) The quantum-critical state is essentially identical to the two-channel Kondo state;
(ii) Particle-hole or device (parity) symmetry are {\em not} required;
(iii) The critical point is destroyed once there is charge transfer between
channels 1 and 2.
The phase diagram obtained from these simple arguments is shown in Fig.~\ref{fig:phase}.
A similar picture is obtained within a CFT approach \cite{ALJ}.

{\em Conformal field theory.}
Since we do not have electron-hole symmetry in any of the channels,
{\pf
we used only the  symmetries $U_1(1)$  and $U_2(1)$ associated with
charge conservation in the two channels and the global spin $SU(2)_2$ symmetry
for the conformal field theory solution.
In the corresponding coset construction, $U_1(1)\times U_2(1) \times SU(2)_2
\times \Zed_2$ \cite{ALJ}, all primary states and
primary fields are characterized by their two charges, $Q_1$ and $Q_2$, their
spin $j$, and an Ising quantum number $q$ ($Id$, $\sigma$, and $\epsilon$).}

{\pf At the critical point the entire finite-size spectrum can be characterized just by two
phase shifts,  $\delta_1,\delta_2\in [0,\pi/2]$ }. Similar to Ref.~\onlinecite{ALJ},
the finite size spectrum is obtained through fusion with the Ising field $\sigma$.
The leading relevant operators at the fixed point are listed in Table~\ref{table}, where
we also indicated the total charge, $Q=Q_1+Q_2$ of every operator.
Only operators with $Q=0$ can occur in the Hamiltonian, and in the absence
of magnetic field only spinless operators can appear, therefore
there are only {\em two} possible relevant operators,
$\phi$ and $\phi_\pm$ that can be present in the Hamiltonian.
Therefore, in the vicinity of the QCP,
the Hamiltonian can be written as
\be
H = H^* + \kappa \;\phi + \delta \;  \phi_+  +  \delta^* \; \phi_- \dots,
\label{eq:H_fp}
\ee
where $H^*$ denotes the fixed-point Hamiltonian.
The coefficient $\kappa \approx \delta K/\sqrt{T_K} = (K-K_C)/\sqrt{T_K}$
measures the distance to the critical point,
with $T_K\sim \min\{ T_1,T_2\}$ being the Kondo scale associated with the formation of
the non-Fermi liquid state.
From the quantum numbers it is clear that the operators  $\phi_\pm$ transfer exactly one
charge from one side to the other, therefore the coefficient of
$\delta$ is related to the amplitude of those operators in
Eq.~(\ref{eq:H_gen}) that transfer charge between the two sides, and which have been
neglected in Eq.~(\ref{eq:H_simple}).
Both operators have scaling dimension $1/2$~\cite{chargefoot},
and are thus relevant at the fixed point.
However, $\kappa$ can be tuned to zero, while $\delta$ always takes a finite value and
generates a smooth cross-over at an energy scale $T_\delta^*\sim |\delta|^2$,
even for $\kappa=0$.
As a result, a double-dot system never displays a true QPT.
Nevertheless, as we shall see later, the parameter $\delta$ can be made small in a
controlled way, such that the structure of the quantum critical point $\kappa = \delta=0$
can be explored.

\begin{table}
\begin{tabular}{cccccc}
\hline
\hline
 $ Q$  & $(Q_1,Q_2)$  & $j$ & Ising & $x$ & operator
\\
\hline
 0 & (0,0) & 0 & $Id$ & 0  & $\sim H^*$
\\
 0 & (0,0) & 0 & $\epsilon$ & $1\over 2$ & $ \phi \sim \delta K$
\\
 0 & $\pm(1,-1)$ & 0 & $Id$ & ${1\over 2} \mp \frac{\delta_1 - \delta_2}\pi$
& $\phi_\pm \sim \psi^\dagger_1 \psi_2$, $\psi^\dagger_2 \psi_1$
\\
 0 & (0,0) & 1 & $Id$ & $1\over 2$ & $ \vec \phi \sim \vec B$
\\
$\pm1$  & $\pm (1,0)$ & 1/2 & $\sigma$ & $\frac 12  \mp \frac{\delta_1}\pi $ &  $\sim \psi^\dagger_1,\psi_1$
\\
$\pm1$  & $\pm (0,1)$ & 1/2 & $\sigma$ & $\frac 12  \mp \frac{\delta_2}\pi $ &  $\sim \psi^\dagger_2,\psi_2$
\\
$\pm2$  & $\pm (1,1)$ & 0 & $Id$ & $\frac 12  \mp \frac{2(\delta_1+\delta_2)}\pi $ & ~~ $\sim \psi^\dagger_1 i\sigma_y \psi^\dagger_2$,{\pf $\psi_1 i\sigma_y \psi_2$ }
\\
\hline
\hline
\end{tabular}
\caption{
\label{table}
Operator content of the critical point.
}
\end{table}


{\em Renormalization group.}
To obtain an estimate for the (dangerous) coupling $\delta$ in Eq.~(\ref{eq:H_fp})
we need to compute the renormalization of the various processes that
correspond to charge transfer in Eq.~(\ref{eq:H_gen}) \cite{Varma_unpub}.
To this purpose, let us assume that $T_1\approx T_2\approx T_K$
and construct the perturbative scaling equations
for the couplings in Eq.~(\ref{eq:H_gen}).
In leading logarithmic order they read
\be
\frac{d{\underline J^{(\gamma)}}}{dl} = \bigl(\underline J^{(\gamma)}\bigr)^2\;, ~~
\frac{d{\underline Q^{(\gamma)}}}{dl} = \frac{d{\underline V^{(\gamma)}}}{dl} = 0\;.
\ee
Here $l = \ln(\delta\epsilon/\Lambda)$ denotes the logarithmic energy scale, and
we introduced a matrix notation in the lead indices,
$Q_{\gamma\gamma'} \to \underline Q,\dots$. From these equations we readily see that
the most dangerous operators are the off-diagonal parts of the $J^{(\gamma)}$ which
increase along the RG flow.
However, in the perturbative regime the ratios $J^{(1)}_{12}/J_1$ and $J^{(2)}_{12}/J_2$
remain approximately constant.
At the scale $T_K$ we have $J_1\sim J_2\sim 1$, from which we immediately obtain an
estimate for the parameter $\delta$: $\delta \sim  \sqrt{T_K}  (K/E_C)^{1/2}$.
Thus, for a double-dot system we find:
$
T^*_{\delta,DD} \sim T_K K/{E_C}\;.
$
For typical semiconductor quantum-dot parameters,
$E_C\sim 20\; {\rm K}$, and $K\sim T_K \sim 0.5 \;{\rm K}$,
this gives a cross-over scale $T_{\delta,DD}^* \sim 12 \;{\rm mK}$,
which, while not very large, might be enough to spoil an observation of the non-Fermi
liquid behavior.


{\em Suppressing charge transfer.}
$T_{\delta}^*$ can be suppressed by creating an artificial antiferromagnetic
insulator to mediate the exchange interaction between the two main dots 1,2.
The simplest arrangement is shown in Fig.~\ref{fig:DD}b,
where we connect the two dots with two additional quantum dots with one electron on each of them.
For simplicity, let us assume that the charging energies of all dots are similar, but
the tunneling-generated exchange coupling $K_2$ between the two central dots
is somewhat larger than the one between the outer dots and their neighbors,
$E_C>K_2> K_1$ (see Fig.~\ref{fig:DD}).
In this limit, at energy scales below $K_2$ the spins on the central dots
are bound to a singlet, and their role is essentially restricted
to mediate an antiferromagnetic interaction $K \sim K_1^2/K_2$
between the two main dots.
With parameters $K_2\approx 3 \;{\rm K}$ and $K_1 \approx  1.5\; {\rm K}$
this gives a coupling in the range of $K\sim 1 {\rm K}\sim T_K$.
On the other hand,  $J^{(\gamma)}_{12}\sim (J_1 J_2 K_2 K_1^2 /E_C^3)^{1/2}$,
and therefore $T_\delta^*$ is reduced to
\be
T^*_{\delta,4D} \sim T_K \left(\frac{K_1}{E_C}\right)^2 \frac{K_2}{E_C}\;.
\ee
With the above parameters we find $T_{\delta,4D}^* \approx 10^{-3}\; T_K \approx  0.5 \;{\rm mK}$.
This value can readily be decreased even further by inserting more quantum dots in the
middle.


{\em Transport.}
In the remainder of the paper we thus assume that $T_\delta^*$ is smaller
than the experimentally relevant temperature scales, i.e., we set $\delta=0$.
Let us furthermore concentrate on $T_1\approx T_2\approx T_K$.
CFT allows to predict various observables in the regime close to the QCP, $\kappa\approx0$.
We first note that in the absence of charge transfer, the linear conductance through dot
$\gamma$ is simply related to the $T$-matrix $T^{(\gamma)}$ of the conduction
electrons in the corresponding electrodes as $G_1 =G^{(1)}_0 \;{\rm Im}\{ T^{(1)}/2\}$ with
$G^{(1)}_0 = \frac{2e^2}h\; 4 \Gamma_{L1} \Gamma_{R1} /(\Gamma_{L1} +\Gamma_{R1} )^2$
(see Fig.~\ref{fig:DD}).
At the fixed point (i.e., zero temperature),
$T^{(1)} = i(1-S^{(1)})$, with  $S^{(\gamma)}$ the $S$-matrix of the electrons in
lead $\gamma$ \cite{AL_2CK_PRB}. Similar to the analysis of  \cite{AL_2CK_PRB}
we find that $S^{(1)}=S^{(2)}=0$ at the QCP,
and thus the conductance is  $G_1(T=0) = G^{(1)}_0/2$ for $K=K_C$.
The approach to this value is determined by the leading irrelevant operator,
which, similar to the electron-hole symmetrical case, is $\phi'$, the derived field
from $\phi$~\cite{asymfoot}.
At $K = K_C$, the finite-temperature corrections to $G_0^{(1)}$ can be computed by
perturbation theory in $\phi'$, with the result
$G_{1,QCP}(T) = G^{(1)}_0 \left(1 - \alpha_1 \sqrt{T/T_K}+\dots\right)$.
Here $\alpha_1$ is a non-universal constant of order unity that depends on the
asymmetry of the device and on the phase shifts.
At finite source-drain voltages, $V$, the deviation  $\delta G_1 \equiv G^{(1)}_0 - G_1(T)$
will display scaling properties, similar to those of the 2CK model~\cite{Buhrman,Kroha}
\be
{\delta G_1}/{G_0}  = \sqrt{T/T_K} \; F(V/T)\;,
\ee
where the (non-universal) function $F$ has the properties
$F(x\ll1)\approx {\rm const}$ and $F(x\gg 1)\propto \sqrt{x}$.

For small but finite $\kappa$, another crossover occurs
at an energy scale $T_\kappa^* = \kappa^2 \approx (K-K_C)^2/T_K$:
For $\kappa>0$ a inter-impurity singlet state is formed,
while for $\kappa<0$ a Kondo state is recovered.
At these fixed points the $S$-matrices are given by
$S^{(\gamma)}= e^{2i \delta_{\gamma}}$ ($K>K_c$) and $S^{(\gamma)}= -e^{2i\; \delta_{\gamma}}$
($K<K_C$), with
both of these fixed points are of Fermi-liquid type, and therefore
the conductance at them scales as
$G_{1,{\rm singlet}} = G_0\; \bigl(\sin^2(\delta_1) + \beta_1 (T/T_\kappa^*)^2+\dots\bigr) $
and $G_{1,{\rm screened}} = G_0 \;(\cos^2(\delta_1) - \gamma_1 (T/T_\kappa^*)^2+\dots) $,
respectively, with $\beta_1$ and $\gamma_1$ again non-universal constants
of order of unity \cite{phasefoot}.
The properties of $G_1(T)$ are summarized in  Fig.~\ref{fig:phase}.

\begin{figure}
\includegraphics[width=7 cm]{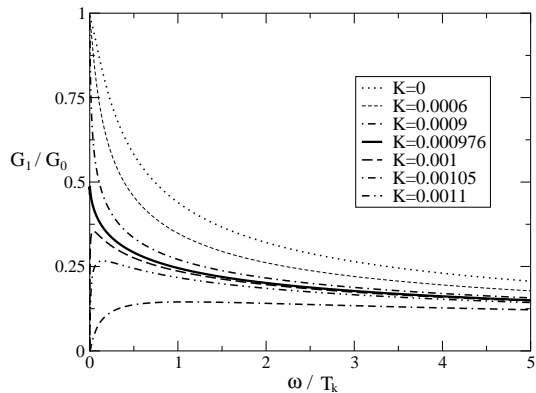}
\caption{\label{fig:NRG}
NRG results for $G_1(\omega)$ (in unit of $G_0$)
for different RKKY couplings. The parameters used here correspond
to Kondo couplings $J_1\approx 0.1$,
$J_2\approx 0.2$, the potential scattering terms $V_1\approx 0.003$,
$V_2\approx 0.02$, where the energy unit is the half bandwidth
of the conduction electrons.
The critical RKKY coupling is $K_c\approx 0.000976$. The frequency $\omega$
in the plot is in units of $T_K$ where $T_K$ is defined as the half-width of
$G_1(\omega)$ for $K=K_c$.
}
\vspace*{-5pt}
\end{figure}

A numerical computation of the finite-temperature scaling functions
in the vicinity of the QCP is notoriously difficult.
However, we can compute the AC conductance $G_1(\omega)$ \cite{Hofstetter}
by applying the NRG approach to the Anderson Hamiltonian corresponding to
Eq.~(\ref{eq:H_simple}).
The results of this calculation for a generic situation without particle-hole and
parity symmetries are shown in Fig.~\ref{fig:NRG}.
The various crossovers can be clearly observed in $G_1(\omega)$ as a function
of frequency, which displays a behavior qualitatively similar to $G_1(T)$.


{\em Summary.}
We have demonstrated that the quantum phase transition of the two-impurity
Kondo model can be experimentally accessed using double quantum-dot devices.
The non-Fermi liquid state is robust against particle-hole and device
asymmetries; it is destroyed by charge transfer between the two main dots,
which, however, can be effectively suppressed with additional quantum dots in
the set-up. Using a combination of analytical and numerical methods we have
made predictions for relevant energy scales and transport quantities.

We thank C. M. Varma for valuable discussions.
This research was supported by the Hungarian Grants
OTKA Nos. NF061726, T046267, and T046303, and
by the DFG Center for Functional
Nanostructures Karlsruhe.

\vspace*{-15pt}
\bibliographystyle{apsrev}

\begin{thebibliography}{30}
\bibitem{renormfoot}
Here we neglected the renormalization of the density of states
produced by $V_1$ and $V_2$.
\bibitem{asymfoot}
For asymmetric setups, unlike in Ref.~\onlinecite{ALJ},
the operator $\vec J_{-1} \vec \phi$ is also allowed.
\bibitem{phasefoot}
The phase shifts appearing in these formulas may be different
for the singlet and Kondo screened fixed points.
\end{thebibliography}

\end{document}